\documentclass[twocolumn,pra,aps,showpacs]{revtex4}

\usepackage{mathptmx}
\usepackage{subfigure}
\usepackage{psfrag,graphicx}
\usepackage{dcolumn}
\usepackage{amsmath,amssymb}
\usepackage{bm}
\usepackage{color}
\usepackage{latexsym}
\usepackage{epstopdf}
\usepackage{color}
\usepackage[english]{babel}
\usepackage{latexsym}
\usepackage{psfrag,graphicx}
\usepackage{subfigure}
\usepackage{amsmath}
\usepackage{amssymb}
\usepackage{amsfonts}
\usepackage{bm}
\usepackage{natbib}
\usepackage{epstopdf}
\usepackage{appendix}

\definecolor{mygrey}{gray}{0.35}
\definecolor{myblue}{rgb}{0.2,0.2,0.8}
\definecolor{myzard}{cmyk}{0,0,0.05,0}
\definecolor{mywhite}{rgb}{1,1,1}
\definecolor{mywhite}{rgb}{1,1,1}
\definecolor{myred}{rgb}{1,0.,0.3}

\usepackage[colorlinks=true,citecolor=myblue,linkcolor=myred]{hyperref}

\def\ba{\begin{align}}
\def\enda{\end{align}}
\def\bi{\begin{itemize}}
\def\ei{\end{itemize}}

\def\be{\begin{equation}}
\def\ee{\end{equation}}
\def\bea{\begin{eqnarray}}
\def\eea{\end{eqnarray}}
\def\bse{\begin{subequations}}
\def\ese{\end{subequations}}

\begin{document}
\title{Steady-State Force Sensing with Single Trapped Ion}
\author{Peter A. Ivanov}
\affiliation{Department of Physics, St. Kliment Ohridski University of Sofia, James Bourchier 5 blvd, 1164 Sofia, Bulgaria}

\begin{abstract}
We propose a scheme for detecting time-varying weak forces using quantum probe consisting of single spin and quantum oscillator under the effect of collective dissipation. We study the force estimation in the steady-state regime where the information of the force is extracted by measuring observable of the oscillator such as quadrature and mean phonon excitation. We quantify the force sensitivity in terms of quantum Fisher information and show that it diverges approaching the critical spin-boson coupling making the system sensitive to very small force perturbation. We show that close to the critical coupling the measurement of the oscillator quadrature is optimal in a sense that saturates the fundamental Cramer-Rao bound. Furthermore, we study the force estimation in the presence of phonon squeezing and show that it can significantly improve the sensitivity reaching minimal detectable force of order of xN (1${\rm xN=10^{-27}{\rm N}}$).

\end{abstract}

\maketitle

\section{Introduction}
One of the most promising avenue of the quantum technologies is the high precise measurements of very weak forces which has broad application in quantum metrology and quantum sensing \cite{Giovannetti2011,Giovannetti2006,Degen2017} as well as for fundamental tests of gravity \cite{Geraci2008,Arvanitaki2013}. Usually, the force detector relies on using mechanical oscillator as a quantum probe sensitivity to very small displacement \cite{Butt2005}. Examples include optomechanical systems where force sensitivity in the range of zN (1${\rm zN}=10^{-21}{\rm N}$) has been demonstrated \cite{Teufel2009,Moser2013,Kampel2017}. Trapped ion system is another example of highly sensitive system to a small displacement where the vibration degree of freedom plays a role of quantum mechanical oscillator. High precision control over the motion and internal spin states of the trapped ion allows to reach force sensitivity in the range of yN (1${\rm yN}=10^{-24}{\rm N}$) \cite{Maiwald2009,Biercuk2010,Gloger2015,Shaniv2017}. Recently, an amplitude sensing below the zero-point fluctuation was demonstrated with ions in a Penning trap \cite{Gilmore2017}. Further improvement of the force sensitivity can be achieved by squeezing the motional mode of the trapped ion which leads to amplification of mechanical oscillator displacements \cite{Burd2018}.

In this work we propose force sensing in the presence of driven dissipative processes which causes losses of excitations of the quantum oscillator. Our probe consists of a single harmonic oscillator coupled coherently with an effective spin system via dipolar coupling described by the quantum Rabi (QR) model. Such a model can be implemented in a various quantum optical systems including for example photonic \cite{Crespi2012}, superconducting circuits \cite{Mezzacapo2014} and trapped ion systems \cite{Lv2018}. In particular, the trapped ion realization relies on using the laser induced coupling between the vibrational degree of freedom and the electronic states of the ion \cite{Schneider2012}. Recently, quantum sensing scheme for the estimation of very weak force based on the coherent dynamics of the QR model was proposed \cite{Ivanov2015,Ivanov2016,Ivanov2016_1}. Here, we extend the discussion by considering the effect of the dissipation during the force estimation. In our scheme the information of the force is extracted by measuring experimental observable such as the quadratures and mean phonon number of the quantum oscillator after the system approaches the steady state. We consider the strong coupling regime in which the spin states can be adiabatically eliminated, which causes squeezing of the motional degree of freedom. We show that this effect improves the signal-to-noise ratio even in the presence of dissipation such that the force sensitivity can be enhanced by increasing for example the spin-phonon coupling. We quantify the force sensitivity in terms of quantum Fisher information and show that diverges by approaching the critical point making the system sensitive to very small force perturbation. Furthermore, we propose force estimation in the presence of parametric amplification. In that case the quantum probe is described by QR model with additional squeezing term. We follow the adiabatic sensing technique \cite{Ivanov2015} and show that by using phonon squeezing the force sensitivity can be improved by order of magnitude better compared to a non-squeezing force estimation. We show that by using squeezing of order of few kHz the minimal detectable force extracted by measuring the spin state populations can reach regime of order of xN (1${\rm xN=10^{-27}{\rm N}}$).

The paper is organized as follows. In Sec. \ref{model} we introduce the quantum probe sensitivity to time-varying external force, which consists of a single harmonic oscillator interacting with the effective spin states via dipolar coupling. Additionally, the dissipation of the quantum harmonic oscillator excitations is included. In Sec. \ref{SSFS} we discuss the steady state solution of the model and show that the force estimation can be performed by measuring experimental observable of the oscillator such as quadratures or the mean phonon number. In Sec. \ref{Sq} we discuss the improvement of the force sensitivity by using squeezing of the motion-degree of freedom. Finally, in Sec. \ref{C} we summarize our findings.

\section{Model}\label{model}
The quantum system of interest consists of a single bosonic mode and an effective spin system which interact via dipolar coupling. Such a system is described by the Hamiltonian
\begin{eqnarray}
&&\hat{H}=\hat{H}_{0}+\hat{H}_{\rm sb}+\hat{H}_{\rm F},\quad \hat{H}_{0}=\hbar\omega \hat{a}^{\dag}\hat{a}+\frac{\hbar\Omega}{2}\sigma_{x},\notag\\
&&\hat{H}_{\rm sb}=\hbar g(\hat{a}^{\dag}+\hat{a})\sigma_{z},\quad \hat{H}_{\rm F}=\frac{zF}{2}(\hat{a}^{\dag}+\hat{a}),\label{Rabi}
\end{eqnarray}
where $\hat{a}^{\dag}$ and $\hat{a}$ are the creation and annihilation operators corresponding to an quantum harmonic oscillator with frequency $\omega$ and $\sigma_{x,y,z}$ is the respective Pauli matrix. $\Omega$ and $g$ are the transverse field and spin-boson coupling, respectively. Additionally, an external driving force is applied which displaces the motional amplitude of the oscillator described by the term $\hat{H}_{\rm F}$, where $z$ is the spread of the zero-point wavefunction and $F$ is the parameter we wish to estimate. In the absence of the driving term $\hat{H}_{\rm F}$ the Hamiltonian resemble the quantum Rabi model (QR), which possesses a discrete symmetry defined by transformation $\sigma_{y,z}\rightarrow -\sigma_{y,z}$ and $\hat{a}\rightarrow-\hat{a}$ which implies that $\hat{H}\rightarrow \hat{H}$. There are various quantum optical systems where QR model can be realized \cite{Niemczyk2010}. For example, using trapped ion the model is implemented by applying bichromatic laser fields with laser frequencies $\omega_{{\rm L},{\rm b}}=\omega_{0}+\omega_{\alpha}-\omega$ and $\omega_{{\rm L},{\rm r}}=\omega_{0}-\omega_{\alpha}+\omega$ which addresses the vibrational blue- and red-sidebands \cite{Pedernales2015,Lv2018}. Here $\omega_{0}$ is the transition frequency between the two metastable ion states $|s\rangle$ ($s=\uparrow,\downarrow$), $\omega_{\alpha}$ is the trap frequency and $\omega$ is the detuning. We assume that external time-varying force is applied $F(t)=F\cos((\omega_{\alpha}-\omega)t)$ which displaces the ion's motion with Hamiltonian $\hat{H}_{\rm F}(t)=z F(t)(\hat{a}^{\dag}+\hat{a})$. Performing optical and vibrational rotating-wave approximation we arrive in the QR Hamiltonian (\ref{Rabi}) \cite{Wineland1998}.
\begin{figure}[tb]
  \includegraphics[width=0.9\columnwidth]{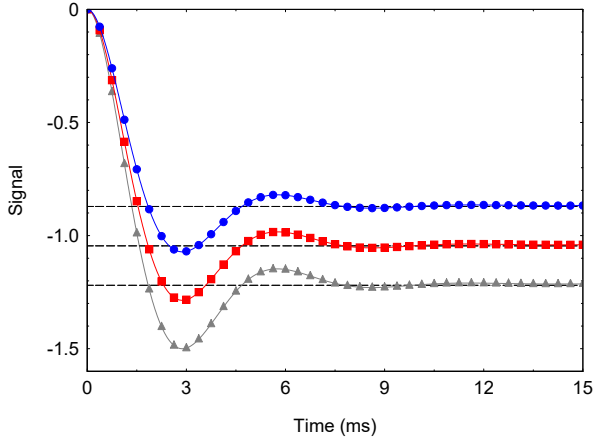}
  \caption{Expectation value of the position operator $\hat{x}$ versus the interaction time. The parameters as set to $\Omega/2\pi=320$ kHz, $g/2\pi=4.0$ kHz, $\omega/2\pi=0.3$ kHz, $\gamma/2\pi=0.08$ kHz, $z=14$ nm. The numerical solution of the master equation (\ref{master}) with Hamiltonian (\ref{Rabi}) for $F=5.0$ yN (blue dots), $F=6.0$ yN (red squares), $F=7.0$ yN (grey triangles) is compared with the asymptotic steady state formula (\ref{x}) (dashed lines).}
  \label{signal}
\end{figure}

Within the framework of master equation of Lindblad form the dynamics of the system is described by
\begin{equation}
\partial_{t}\hat{\rho}=-\frac{i}{\hbar}[\hat{H},\hat{\rho}]+\hat{\mathcal{L}}(\hat{\rho}),\label{master}
\end{equation}
where $\hat{\rho}$ is the density operator of the system. The term $\hat{\mathcal{L}}(\rho)$ describes the dissipative dynamics of the bosonic mode. Hereafter we consider bosonic decay described by the dissipative term
\begin{eqnarray}
\hat{\mathcal{L}}(\hat{\rho})&=&\gamma(2\hat{a}\hat{\rho}\hat{a}^{\dag}-\{\hat{a}^{\dag}\hat{a},\hat{\rho}\}),\label{L}
\end{eqnarray}
where $\gamma$ is the decay rate which we assume to be positive. The interplay between the coherent and incoherent dynamics gives rise to a steady state solution of the master equation (\ref{master}) determined by the condition $\partial_{t}\hat{\rho}=0$.
\begin{figure}[tb]
  \includegraphics[width=1.0\columnwidth]{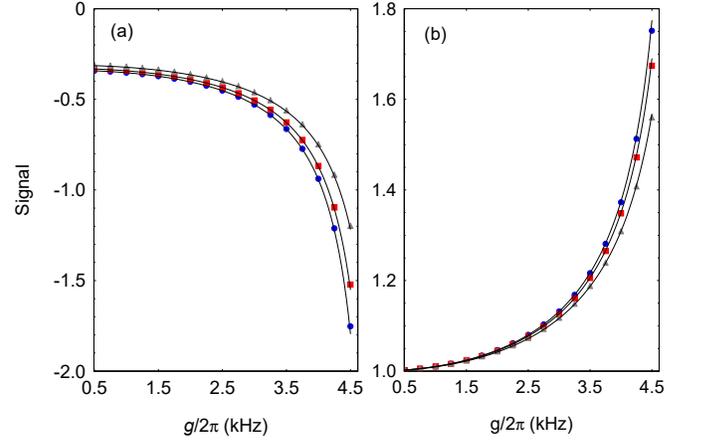}
  \caption{(a) Expectation value of the position operator $\hat{x}$ versus the spin-boson coupling $g$ for different frequencies $\omega$. We compare the numerical solution of the master equation (\ref{master}) with Hamiltonian (\ref{Rabi}) for $\omega/2\pi=0.29$ kHz (blue circles), $\omega/2\pi=0.30$ kHz (red squares), $\omega/2\pi=0.32$ kHz (grey triangles). The solid line is the analytical result (\ref{x}). (b) Variance of the signal $\langle\Delta \hat{x}\rangle$. The solid line is Eq. (\ref{var_x}).}
  \label{Fig2}
\end{figure}

In the following we are interested in the time-evolution of the expectation value of the hermitian operator $\hat{A}$ which is governed by the Heisenberg equation of motion, namely
\begin{equation}
\partial_{t}\langle \hat{A}\rangle={\rm Tr}(\hat{A}\partial_{t}\hat{\rho}).
\end{equation}
By approaching the steady-state regime the expectation values can be determined by the condition $\partial_{t}\langle \hat{A}\rangle=0$. The force sensing scheme consists of measuring the observable $\hat{A}$ in the steady state, where the shot-noise limited sensitivity in the estimation of the force $F$ from the measured signal $\langle \hat{A}\rangle$ is given by
\begin{equation}
\delta F_{A}=\frac{\langle\Delta \hat{A}\rangle}{\sqrt{\nu}\frac{\partial \langle \hat{A}\rangle}{\partial F}},\label{SNR}
\end{equation}
where $\langle \Delta\hat{A}\rangle=\sqrt{\langle \hat{A}^{2}\rangle-\langle \hat{A}\rangle^{2}}$ is the variance of the signal and $\nu=T/\tau$ is the experimental repetition number. Here $T$ is the total experimental time and $\tau$ includes the evolution, preparation and measurement time.

\section{Steady State Force Sensing}\label{SSFS}

In order to perform force measurement by detecting the bosonic degree of freedom of the system we consider strong coupling regime $g\gg\omega$. Assuming also that transverse field is much larger than all other energy scale of the system $\Omega\gg\omega,g$ the spin-degree of freedom becomes frozen such that they can be traced out which leads to a pure bosonic model. Such a limit of the QR model was considered in the context of quantum phase transition in closed as well as in open systems \cite{Hwang2015,Hwang2018}. The adiabatic elimination of spin degree of freedom can be achieved by making unitary transformation of the density operator $\rho=\hat{R}\tilde{\rho}\hat{R}^{\dag}$ where $\hat{R}=e^{\hat{S}}$ with $\hat{S}$ being anti-Hermitian operator. Inserting the latter in Eq. (\ref{master}) we obtain the effective Hamiltonian $\hat{H}_{\rm eff}=\hat{R}^{\dag}\hat{H}\hat{R}$. We choose the operator $\hat{S}$ such that all terms in order of the coupling $g$ are canceled and the first term describing the spin boson interaction is of order of $g^{2}/\Omega$. We have $\hat{S}=\frac{ig}{\Omega}\sigma_{x}(\hat{a}^{\dag}+\hat{a})$ and the effective Hamiltonian becomes $\hat{H}_{\rm eff}=\hat{H}_{0}+\frac{1}{2}[\hat{H}_{\rm sb},\hat{S}]+\hat{H}^{\prime}$, where $\hat{H}^{\prime}=\frac{1}{2}[[\hat{H}_{\rm sb},\hat{S}],\hat{S}]+\ldots$ contain terms of order of $O(g^{3}/\Omega^{2})$ which we neglect in the limit $\Omega\gg g$. Moreover, in this limit the dissipative term in (\ref{master}) is not affected by the transformation such that we have $\hat{R}^{\dag}\hat{\mathcal{L}}(\hat{\rho})\hat{R}\approx \hat{\mathcal{L}}(\hat{\rho})$. Indeed, the unitary transformation $\hat{R}$ gives rise to a spin-dependent displacement of the bosonic mode with amplitude proportional to $g/\Omega$ which can be neglected as long as $\Omega\gg g$. The effective Hamiltonian becomes
\begin{equation}
\hat{H}_{\rm eff}=\hbar\omega\left(1-\frac{\lambda^{2}}{2}\right)\hat{a}^{\dag}\hat{a}-\hbar\frac{\omega \lambda^{2}}{4}(a^{\dag2}+\hat{a}^{2})+\frac{z F}{2}(\hat{a}^{\dag}+\hat{a}),\label{Heff}
\end{equation}
where we have assumed that the spin is initially prepared in the state $\left|-\right\rangle$ ($\sigma_{x}|\pm\rangle=\pm|\pm\rangle$) and $\lambda=2 g/\sqrt{\omega\Omega}$.

In the following we discuss the steady-state expectation values of the experimental observable which can be used to estimate the force, namely the quadratures and the mean phonon number of the quantum oscillator.

\subsection{Force estimation by measuring quadratures}
\begin{figure}[tb]
  \includegraphics[width=1.0\columnwidth]{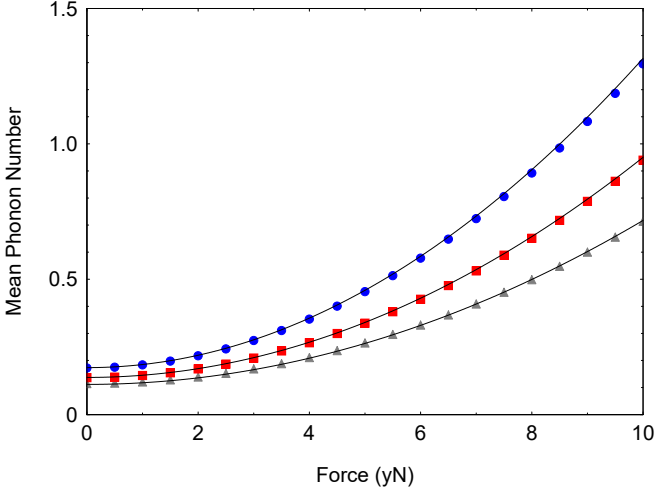}
  \caption{Mean phonon number versus the force $F$ for different frequencies $\omega$. We compare the numerical solution of the master equation (\ref{master}) with Hamiltonian (\ref{Rabi}) for $\omega/2\pi=0.32$ kHz (grey triangles), $\omega/2\pi=0.30$ kHz (red squares), and $\omega/2\pi=0.28$ kHz (blue circles). The solid line is the analytical expression (\ref{mpn}).}
  \label{Fig3}
\end{figure}
In the absence of force term the parity symmetry of Hamiltonian (\ref{Heff}) is preserved and one can expect that the expectation values of the bosonic operators become $\langle \hat{a}\rangle=\langle \hat{a}^{\dag}\rangle=0$. However, the effect of the force term is to break the parity symmetry such that the expectation value of the position operator $\hat{x}=\hat{a}^{\dag}+\hat{a}$ approaches the steady state which is given by
\begin{equation}
\langle \hat{x}\rangle=-\frac{\tilde{F}}{\lambda^{2}_{\rm c}-\lambda^{2}},\label{x}
\end{equation}
where $\tilde{F}=\frac{z F}{\hbar\omega}$ and $\lambda_{\rm c}=\sqrt{1+\frac{\gamma^{2}}{\omega^{2}}}$. In Fig. \ref{signal} we show the exact evolution of the quadrature $\langle \hat{x}\rangle$ with Hamiltonian (\ref{Rabi}). As can be seen the system approaches the steady state where very good agreement is observed with the analytical expression (\ref{x}). In Fig. \ref{Fig2}(a) is shown the signal $\langle \hat{x}\rangle$ when the spin boson coupling is varied. Approaching the critical value $\lambda_{\rm c}$ the signal is enhanced and diverges in the limit $\lambda\rightarrow \lambda_{\rm c}$ as $\langle \hat{x}\rangle\sim (\lambda_{\rm c}-\lambda)^{-1}$.
We find that the variance of the position quadrature is independent of the force term $F$ and is given by
\begin{equation}
\langle \Delta\hat{x}\rangle=\sqrt{\frac{(2 \lambda^{2}_{\rm c}-\lambda^{2})}{2(\lambda_{\rm c}^{2}-\lambda^{2})}}.\label{var_x}
\end{equation}
\begin{figure}[tb]
  \includegraphics[width=1.0\columnwidth]{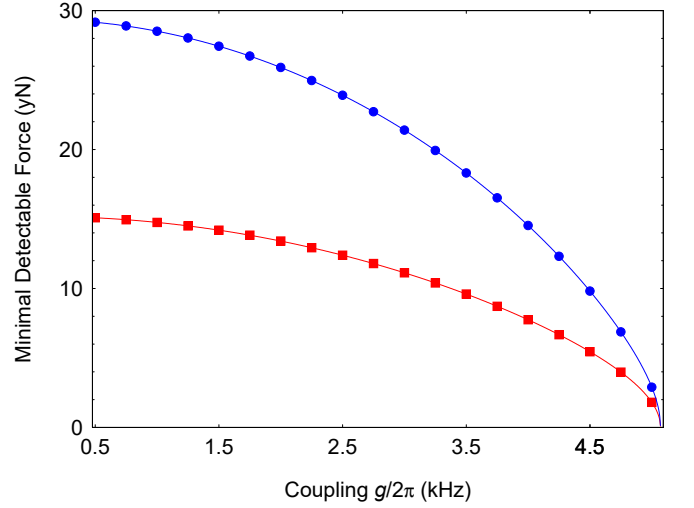}
  \caption{Minimal detectable force versus the coupling strength $g$. The red squares correspond to the observable $\hat{A}=\hat{x}$ and respectively the blue circles to $\hat{A}=\hat{n}$. The parameters are set to $\omega/2\pi=0.30$ kHz, $\Omega/2\pi=320$ kHz, $\gamma/2\pi=0.08$ kHz, $z=14$ nm.}
  \label{Fig4}
\end{figure}
Crucially the variance of the signal diverges as $\langle \Delta \hat{x}\rangle\sim (\lambda_{\rm c}-\lambda)^{-1/2}$ such that the minimal detectable force becomes
\begin{equation}
\delta F_{x}=\frac{\hbar\omega}{\sqrt{2}z}\sqrt{(2\lambda^{2}_{\rm c}-\lambda^{2})(\lambda^{2}_{\rm c}-\lambda^{2})}.\label{Fx}
\end{equation}
We observe that from one hand the force sensitivity decreases for high dissipative coupling $\gamma$ and thus $\lambda_{\rm c}$ but on the other hand it can be improved by increasing the coupling $\lambda$. This can be achieved for example by increasing the spin-boson coupling strength $g$ as is shown in Fig. \ref{Fig2}.

For completeness we provide also the force estimation by measuring the expectation value of the momentum quadrature $\hat{p}=i(\hat{a}^{\dag}-\hat{a})$ and the momentum variance $\Delta \hat{p}$. We find
\begin{equation}
\langle \hat{p}\rangle=-\frac{\tilde{F}\gamma}{\omega}\frac{1}{\lambda_{\rm c}^{2}-\lambda^{2}},\quad \langle\Delta \hat{p}\rangle=\sqrt{\frac{2\lambda_{\rm c}^{2}-3\lambda^{2}+\lambda^{4}}{2(\lambda_{\rm c}^{2}-\lambda^{2})}}.\label{momentum}
\end{equation}
Using (\ref{SNR}) the minimal detectable force becomes
\begin{equation}
\delta F_{p}=\frac{\hbar\omega^{2}}{\sqrt{2}z \gamma}\sqrt{(2\lambda_{\rm c}^{2}-3\lambda^{2}+\lambda^{4})(\lambda_{\rm c}^{2}-\lambda^{2})}.\label{Fp}
\end{equation}
We observe that for $\omega\gg\gamma$ the measurement of the position quadrature provides better estimation, $\delta F_{p}>\delta F_{x}$ while for strong dissipative coupling $\omega\ll\gamma$ we find $\delta F_{p}<\delta F_{x}$. Approaching the critical coupling $\lambda_{\rm c}$ the minimal detectable force scales as $\delta F_{x,p}\sim \frac{\hbar\omega}{z}\lambda_{\rm c}(\lambda_{\rm c}^{2}-\lambda^{2})^{1/2}$.

Finally, the expectation value of the position quadrature can be extracted by applying red- and blue-sideband laser fields which couple the internal spin states and the motional degree of freedom of the oscillator.

\subsection{Force estimation by measuring the mean phonon number}
Other experimentally convenient observable which can be used to estimate the force is the average number of bosonic excitation, $\langle\hat{A}\rangle=\langle\hat{n}\rangle$. In the steady state we obtain
\begin{equation}
\langle \hat{n}\rangle=\tilde{F}^{2}\frac{\lambda_{\rm c}^{2}}{4(\lambda_{\rm c}^{2}-\lambda^{2})^{2}}
+\frac{\lambda^{4}}{8(\lambda_{\rm c}^{2}-\lambda^{2})},\label{mpn}
\end{equation}
which scales in the limit $\lambda\rightarrow \lambda_{\rm c}$ as $\langle \hat{n}\rangle\sim (\lambda_{\rm c}-\lambda)^{-2}$. In contrast to Eq. (\ref{x}) now the signal shows a quadratic dependence with respect to the force $F$ as is shown in Fig. \ref{Fig3}. For the variance of the signal we find
\begin{eqnarray}
\langle \Delta\hat{n}\rangle^{2}&=&\frac{\tilde{F}^{2}\lambda^{6}}{8(\lambda_{\rm c}^{2}-\lambda^{2})^{3}}+\frac{\lambda^{2}}{32(\lambda_{\rm c}^{2}-\lambda^{2})^{2}}
\{\lambda^{6}+4\tilde{F}^{2}(\lambda^{2}+3)\}\notag\\
&&+\frac{(3\lambda^{4}+4\tilde{F}^{2})}{16(\lambda_{\rm c}^{2}-\lambda^{2})}.
\end{eqnarray}
As a figure of merit for the sensitivity we can use the signal-to-noise ratio ${\rm SNR}=\langle \hat{n}\rangle/\langle\Delta \hat{n}\rangle$ which is larger for better estimation. The minimal detectable force can be determined by the condition $\rm{SNR}=1$. Figure \ref{Fig4} shows the minimal detectable force which can be estimated by measuring the position quadrature or mean phonon number of the oscillator. Consider realistic experimental parameters $g/2\pi=4.5$ kHz, $\omega/2\pi=0.28$ kHz we estimate $\delta F_{x}\approx4.4$ yN and respectively $\delta F_{n}\approx7.8$ yN. Further improvement can be achieved by increasing for example the coupling strength $g$ as is shown in Fig. \ref{Fig4}. However, approaching the critical coupling $\lambda\rightarrow\lambda_{\rm c}$ requires $\Omega/\omega\gg 1$ keeping $\lambda$ finite which limits our sensing protocol.

Next, we provide the ultimate precision bound given by the quantum Fisher information. Such a bound is independent on the measurement and it has pure geometrical meaning in terms of distance between two neighboring states that differ slightly in the value of the parameter $F$.

\subsection{Quantum Fisher Information}

The ultimate precision of the force estimation is bounded by the quantum Cramer-Rao bound
\begin{equation}
\delta F_{\rm Q}\geq\frac{1}{\sqrt{\nu I_{\rm Q}}},\label{qcrb}
\end{equation}
where $I_{\rm Q}$ is the quantum Fisher Information (QFI), which measures the distinguishability of two close quantum states that differ infinitesimally in $F$. Indeed, the QFI can be expressed in terms of the Bures distance between two infinitesimally close quantum states with density matrices $\hat{\rho}_{F}$ and $\hat{\rho}_{F+\epsilon}$ as
\begin{equation}
I_{\rm Q}=4(\partial_{\epsilon}d_{\rm B}(\hat{\rho}_{F},\hat{\rho}_{F+\epsilon}))_{\epsilon=0}^{2}.
\end{equation}
Here $d^{2}_{\rm B}(\hat{\rho}_{1},\hat{\rho}_{2})=2(1-\sqrt{\mathcal{F}(\hat{\rho}_{1},\hat{\rho}_{2})})$ with $\mathcal{F}=\{{\rm Tr}(\sqrt{\hat{\rho}_{1}}\hat{\rho}_{2}\sqrt{\hat{\rho}_{1}})^{1/2}\}^{2}$ being the Uhlmann fidelity between two quantum states. In order to evaluate the QFI we point out that the coherent dynamics described by (\ref{Heff}) as well as the dissipative decay of bosonic excitations desribed by the term (\ref{L}) are Gaussian processes that preserve the Gaussian character of the quantum states. Such states are completely determined by their first two moments \cite{Weedbrook2012}. For this goal let's define the real and symmetric covariant matrix with elements
\begin{equation}
\sigma_{i,j}=\frac{1}{2}\langle \hat{X}_{i}\hat{X}_{j}+\hat{X}_{j}\hat{X}_{i}\rangle-\langle \hat{X}_{i}\rangle\langle \hat{X}_{j}\rangle,\label{cm}
\end{equation}
where $\hat{\bold{X}}^{\rm T}=(\hat{x},\hat{p})$ is two dimensional vector. In the steady state the diagonal elements $\sigma_{11}$ and $\sigma_{22}$ are given by Eqs. (\ref{var_x}) and (\ref{momentum}). For the off-diagonal element we find
\begin{equation}
\sigma_{12}=\frac{\lambda^{2}\gamma}{2\omega(\lambda_{\rm c}^{2}-\lambda^{2})}.
\end{equation}
Using the covariant matrix (\ref{cm}) the QFI can be expressed as \cite{Monras2013,Pinel2013}
\begin{equation}
I_{\rm Q}=\Delta\bold{X}^{\prime {\rm T}}\bold{\sigma}^{-1}\Delta\bold{X}^{\prime},
\end{equation}
with $\Delta\bold{X}^{\prime}=\partial_{F}\langle \bold{X}\rangle$. Here we have used that the the covariance matrix as well as the purity $P=(\det \bold{\sigma})^{-1/2}$ are independent on the force $F$. Finally, we obtain the QFI
\begin{equation}
I_{\rm Q}=\left(\frac{\sqrt{2}z}{\hbar\omega}\right)^{2}\frac{2\lambda_{\rm c}^{2}-\lambda^{2}}{(\lambda_{\rm c}^{2}-\lambda^{2})(4(\lambda_{\rm c}^{2}-\lambda^{2})+\lambda^{4})}.\label{QFI}
\end{equation}
which diverges in the limit $\lambda\rightarrow\lambda_{\rm c}$ making the system sensitive to very small forces. Note that recently efficient parametric estimation close to a dissipative phase transition was proposed in \cite{Lorenzo2017}.

The optimal measurements that saturate the quantum Cramer-Rao bound (\ref{qcrb}) are projective measurements formed by the eigenvectors of the symmetric logarithmic derivative (SLD) operator $\hat{\Lambda}_{F}$ which can be written as \cite{Paris2009}
\begin{equation}
\hat{\Lambda}_{F}=2\sum_{m,n}\frac{\langle\psi_{m}|\partial_{F}\hat{\rho}|\psi_{n}\rangle}{p_{m}+p_{n}}|\psi_{m}\rangle\langle\psi_{n}|.
\end{equation}
The most general form of the single-mode Gaussian state is given by $\hat{\rho}=\sum_{n}p_{n}|\psi_{n}\rangle\langle\psi_{n}|$, where $|\psi_{n}\rangle=\hat{R}(\delta)\hat{D}(\alpha)\hat{S}(\zeta)|n\rangle$ and $p_{n}=N^{n}/(1+N)^{n+1}$ being the thermal state probability with the average number of thermal excitations $N=(1-P)/2P$. Here $\hat{R}(\delta)=e^{i\delta \hat{a}^{\dag}\hat{a}}$ is the rotation operator, $\hat{D}(\alpha)=e^{\alpha \hat{a}^{\dag}-\alpha^{*}\hat{a}}$ is the displacement operator, and $\hat{S}(\zeta)=e^{\frac{\zeta}{2}\hat{a}^{\dag 2}-\frac{\zeta^{*}}{2}\hat{a}^{2}}$ is the squeezing operator with $\zeta=r e^{2i\chi}$. In the steady state we find
\begin{eqnarray}
&&\alpha=\frac{\tilde{F}\lambda_{\rm c}}{2(\lambda_{\rm c}^{2}-\lambda^{2})}, \quad\tan(2\chi+2\delta)=\frac{2\gamma}{\omega(2-\lambda^{2})},\notag\\
&&\tanh2r=\frac{\lambda^{2}}{\sqrt{4(\lambda_{\rm c}^{2}-\lambda^{2})+\lambda^{4}}},\quad \tan\delta=\frac{\gamma}{\omega}.
\end{eqnarray}
Using this one can show that SLD operator becomes
\begin{equation}
\hat{\Lambda}_{F}=2(\partial_{F}\alpha) \hat{R}(\delta)\hat{D}(\alpha)\hat{S}(\zeta)(\beta \hat{a}^{\dag}+\beta^{*}\hat{a})\hat{S}^{\dag}(\zeta)\hat{D}^{\dag}(\alpha)\hat{R}^{\dag}(\delta),
\end{equation}
where $\beta=P(\cosh(r)-e^{2i\chi}\sinh(r))$. Finally, we point out that from Eqs. (\ref{Fx}), (\ref{Fp}) and (\ref{QFI}) it follows that in the vicinity of the critical coupling $\lambda_{\rm c}$ we have $\delta F_{\rm Q}\approx F_{x,p}$ indicating that the quadrature measurement is optimal.

In the following we show that by squeezing the motion degree of freedom one can significantly increase the force sensitivity. Using a squeezing parameter relatively small with respect to $g$ and $\omega$ one can achieve minimal detectable force in order of few xN.

\section{Squeezing Phonons for Enhanced Force Sensing}\label{Sq}
\begin{figure}[tb]
  \includegraphics[width=1.0\columnwidth]{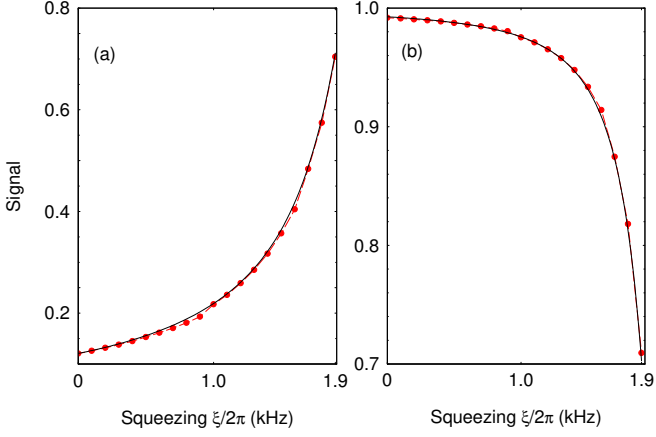}
  \caption{(a) Expectation value of $\sigma_{z}$ at $t_{f}$ as a function of the squeezing parameter $\xi$. We compare the numerical solution with Hamiltonian (\ref{Hsq}) (dashed red circles) with the solution using the two-state approximation (black solid line). The parameters are set to $\omega/2\pi=4.4$ kHz, $g/2\pi=1.6$ kHz, $\Omega_{0}/2\pi=200$ kHz, $F=46$ xN, $\gamma=0$, $\kappa/2\pi=9.5\times 10^{-3}$ kHz, and $t_{f}=284$ ms. (b) Variance of the signal for the same set of parameters.  }
  \label{Fig5}
\end{figure}
Squeezing of motional states has broad range of applications including for example improvement the fidelity of the quantum gates \cite{Ge2019} as well as high precision quantum metrology \cite{Pezze2018}. Recently, highly sensitive detection of small displacement using two orthogonal squeezing operations was demonstrated with single trapped ion \cite{Burd2018}. Here we show that by adding squeezing term to the quantum Rabi Hamiltonian (\ref{Rabi}) one can enhanced the force sensitivity approximately by order of magnitude which allows to reach xN regime of minimal detectable force using squeezing of order of few kHz.

Let's us consider the Hamiltonian
\begin{equation}
\hat{H}=\hat{H}_{\rm R}+\hat{H}_{\rm sq},\quad \hat{H}_{\rm sq}=\hbar\xi(a^{\dag2}e^{i\phi}+\hat{a}^{2}e^{-i\phi}).\label{Hsq}
\end{equation}
Here $\xi$ is the squeezing parameter and $\phi$ is the phase. Hereafter we assume that $\phi=\pi$ which turns out to be optimal for the force detection. Such a squeezing term can be realized by applying an oscillating potential to the trap electrodes with frequency near the twice the motional frequency. We begin by discussing the energy spectrum of the Hamiltonian (\ref{Hsq}) in the limit $\Omega=0$ and setting $\gamma=0$. In that case the model can be treated exactly. Indeed, the Hamiltonian can be written in a diagonal form by making the transformation $\hat{S}^{\dag}(r)\hat{D}^{\dag}(\alpha)\hat{H}\hat{D}(\alpha)\hat{S}(r)=\hbar\omega\sqrt{1-(2\xi/\omega)^{2}}\hat{a}^{\dag}\hat{a}$, where we omit the constant term. Here $\hat{D}(\alpha)$ is the spin-dependent displacement operator with $\alpha=-g\sigma_{z}(\omega-2\xi)^{-1}$ and $\hat{S}(r)$ is the squeezed operator with $r=\frac{1}{4}\ln(1-(2\xi/\omega)^{2})$. The energy spectrum is double degenerate with ground-state doublet given by $|\psi_{\uparrow}\rangle=\hat{D}(\alpha_{\uparrow})\hat{S}(r)\left|\uparrow\rangle|0\right\rangle$ and respectively $|\psi_{\downarrow}\rangle=\hat{D}(\alpha_{\downarrow})\hat{S}(r)\left|\downarrow\right\rangle|0\rangle$, where $\left|n\right\rangle$ ($n=0,1,2\ldots$) is the Fock state of the quantum oscillator. The other set of excited double-degenerate states is $|\psi_{\uparrow,n}\rangle=\hat{D}(\alpha_{\uparrow})\hat{S}(r)\left|\uparrow\rangle|n\right\rangle$ and respectively $|\psi_{\downarrow,n}\rangle=\hat{D}(\alpha_{\downarrow})\hat{S}(r)\left|\downarrow\right\rangle|n\rangle$.

The effect of the transverse field $\Omega$ is to lift the degeneracy of the ground-state manifold, which leads to a effective coupling $\Delta_{\rm c}=\hbar\frac{{\Omega}}{2}\langle\psi_{\downarrow}|\sigma_{x}|\psi_{\uparrow}\rangle$. We follow the adiabatic force sensing scheme proposed in \cite{Ivanov2015} in which the transverse field $\Omega(t)$ varies in time. Preparing the system initially in the paramagnetic phase $|\psi(0)\rangle=|-\rangle|0\rangle$ ($\sigma_{x}|\pm\rangle=\pm|\pm\rangle$) with $\Omega(t_{i})\gg g,\omega,\xi$ the system is adiabatically transferred into the ferromagnetic phase $|\psi(t_{f})\rangle=c_{\uparrow}(t_{f})|\psi_{\uparrow}\rangle+c_{\downarrow}(t_{f})|\psi_{\downarrow}\rangle$ with $\Omega(t_{f})\ll g,\omega,\xi$. In the following we choose $\Omega(t)=\Omega_{0}e^{-\kappa t}$, where $\kappa$ is a characteristic slope. The goal is to estimate the magnitude of the force term during the transition which we treat as a small perturbation by measuring the probabilities $p_{s}(t_{f})=|c_{s}(t_{f})|^{2}$ ($s=\uparrow,\downarrow$). Note that recently, such an adiabatic sweeping of the transverse field was used for the implementation of quantum phase transition between normal/paramagnetic to superradiant/ferromagnetic states in the collective Dicke model \cite{Naini2018}.

Within the two-state approximation the effective Hamiltonian becomes $\hat{H}_{\rm eff}=\Delta_{\rm c}(t)\sigma_{x}+(F_{\uparrow}-F_{\downarrow})\sigma_{z}$, where $F_{\uparrow(\downarrow)}=\langle\psi_{\uparrow(\downarrow)}|\hat{H}_{\rm F}|\psi_{\uparrow(\downarrow)}\rangle$. Using the time-dependence of $\Omega(t)$ the two-state problem is reduced to the the Demkov model \cite{Vitanov1993} where the expectation value of $\sigma_{z}$ is given by
\begin{equation}
\langle\sigma_{z}(t_{f})\rangle=\tanh\left(\frac{\pi g \tilde{F}}{\kappa(1-\frac{2\xi}{\omega})}\right).\label{sigma_z}
\end{equation}
We observe that the effect of the phonon squeezing is to increase the magnitude of the signal to be measured. In Fig. \ref{Fig5}(a) we show the signal as a function of the squeezing parameter where very good agreement is observed between the exact result and the expression (\ref{sigma_z}). The variance of the signal is $\Delta\sigma_{z}=\sqrt{1-\langle\sigma_{z}\rangle^{2}}$ which decreases by increasing $\xi$, see Fig. \ref{Fig5}(b). Using ${\rm SNR}=1$ as a figure of merit for the force sensitivity we find that the minimal detectable force for the parameters in Fig. \ref{Fig5} and squeezing $\xi/2\pi=1.95$ kHz is approximately $F_{\xi,\rm min}=36$ xN. In fact for the same set of parameters but without squeezing the minimal detectable force is $F_{\xi=0,\rm min}=317$ xN such that $F_{\xi=0,\rm min}/F_{\xi,\rm min}\approx 8.7$.  Increasing the squeezing leads to better force estimation but requires longer interaction time since the energy spacing between the ground state manifold and the excited set of states becomes closer. Indeed, the energy difference between the ground and the first excited states is $\Delta E=\hbar\omega\sqrt{1-(2\xi/\omega)^{2}}$ which vanishes as $2\xi$ approaching $\omega$.

Finally, we discuss the effect of the dissipative coupling given by $\hat{\mathcal{L}}(\hat{\rho})$ term in (\ref{master}) on the force estimation. Since the problem is time-dependent no analytical solution can be found for the density matrix. In order to study the effect of the dissipation we integrate numerically the master equation (\ref{master}) with Hamiltonian (\ref{Hsq}). As can be expected the effect of the motional decay is to reduce the force sensitivity. Consider as a example decay rate of $\gamma/2\pi=10^{-3}$ kHz with parameters $\Omega/2\pi=800$ kHz, $\omega/2\pi=5.3$ kHz, $\xi/2\pi=1.0$ kHz and interaction time $t_{f}=120$ ms we find minimal detectable force approximately $F_{\xi,{\rm min}}=1.1$ yN.

\section{Conclusion}\label{C}

We have proposed dissipative estimation protocol for time-varying weak forces using quantum probe which consists of quantum oscillator coupled with effective spin via dipolar coupling. Such a quantum system sensitive to small displacement can be implemented with single trapped ion where the internal electronic states of the ion are coupled to the vibration degree of freedom with bichromatic laser field. Because of the dissipative dynamics the system approaches the steady state which is completely characterized with the first two moments. The force estimation is performed in the steady state by detecting either the quadratures or the mean-phonon number of the oscillator. We have quantified the sensitivity of the force sensing scheme using the quantum Fisher information as a measure for distinguishability of two close quantum states that differ infinitesimally in $F$. We have shown that the quantum Fisher information diverges at the critical spin-phonon coupling making the system sensitive to infinitesimal small force perturbation. We have shown that in the vicinity of the critical coupling the quadrature measurement is optimal in a sense that saturates the fundamental Cramer-Rao bound.

Furthermore, we have discussed the improvement of the force sensitivity using squeezing of vibrational degree of freedom of the oscillator. We have shown that the effect of the phonon squeezing is to enhance the signal to be measured and respectively to reduces the signal variance. We have shown that using experimentally realistic squeezing of few kHz one can significantly improve the sensitivity reaching minimal detectable force in the range of ${\rm xN}$.

\section*{Acknowledgments}

PAI acknowledges support by the ERyQSenS, Bulgarian Science Fund Grant No. DO02/3.


\end{document}